\documentclass[twocolumn,showpacs,preprintnumbers,amsmath,amssym,article]{revtex4}
\usepackage{mathrsfs}
\usepackage{dcolumn}
\usepackage{bm}
\usepackage{graphicx}
\usepackage{MnSymbol,wasysym}
\begin{document}

\title{Coherently controlling robust spin-orbit qubits of electrons in nanowire quantum dots}
\author{Kuo Hai,\footnote{ron.khai@gmail.com} \ Xuefang Deng, \ Qiong Chen, \ Wenhua Hai\footnote{whhai2005@aliyun.com}}
\affiliation{Department of Physics and Key Laboratory of Low-dimensional Quantum Structures and Quantum Control of Ministry
of Education, Hunan Normal University, Changsha 410081, China}
\begin{abstract}
We consider an electron confined in a gated nanowire quantum dot (NQD) with arbitrarily strong spin-orbit coupling (SOC) and weak static magnetic field, and treat the latter as a perturbation to seek the maximal spin-motion entangled states associated with the exact general solutions of the perturbed equations. From the boundedness and self-consistence conditions of the general solutions we find two corrected energies to any $n$ level of the unperturbed system, which have level splitting being much less than the unperturbed level-difference and correspond to a spin-orbit qubit. We demonstrate the metastability of the two-level states and the decoherence-averse effect of SOC, meaning insensitivity of the qubit to the local perturbations and the weak noise from environment. We suggest an alternative scheme to perform the qubit control, simply by adjusting the orientation of magnetic field to produce the quantized phase jumps for any fixed SOC. Such an adjustment can lead to the spin flipping of the state vectors and the position exchanging of the probability-density wavepackets which can be proposed as the non-Abelian quasiparticles. The results could be directly applied to a weakly coupled array of NQDs for coherently encoding the robust spin-orbit qubits.

\end{abstract}


\maketitle

\section{Introduction}

Electrons confined in a nanowire quantum dot (NQD) with spin-orbit coupling (SOC) have been studied and many important results were found \cite{Nowack,NPerge2,Mourik,Ladriere,NPerge3,RLi}. The spin-up and spin-down internal states of the system can be entangled by the external motional states to form a spin-orbit qubit. Phase coherence of the motion states can be used to manipulate the qubit \cite{Leibfried,Monroe,Kitagawa}.
Coherent manipulation of electron spins is one of the central problems of spintronics \cite{Wolf,Loss,Golovach} and is essential for spin-based quantum computing and
information processing \cite{Loss}. The previous investigation has paved the way for manipulating electron spins in nanowire quantum-dot-electron chain individually \cite{Kato,Rashba}. Recently, the search for non-Abelian quasiparticles in semiconducting NQDs with strong SOC has been a focus of theoretical and experimental efforts \cite{Elliott,Mourik,Das,Deng,Ptok,Nilsson,Gazibegovic}, motivated by their potential utility for fault tolerant topological quantum computation \cite{Kitaev,Nayak,Stern,DSau,Stern2,YLi,Else}.
A Majorana particle \cite{Kitaev,LFu, Gazibegovic,XWen} is an electrically neutral non-Abelian anyon \cite{Stern,Abdumalikov,Wilcze} identical to its own antiparticle. Interchanging the Majorana particles changes the state of the system in a way that depends only on the order in which the exchange was performed, which is cornerstone of the braiding operations for encoding topological qubits \cite{Stern}. It has also been demonstrated theoretically \cite{Lutchyn,Oreg} and experimentally \cite{Mourik,NPerge,QHe} that the elusive Majorana particles can be detected in some one-dimensional (1D) systems, including the semiconducting NQD with strong SOC and large $g$ factor \cite{Mourik,Nilsson}, and in proximity to a superconductor. One of the current main objectives may be the investigation of novel alternative models for searching non-Abelian Majorana-like quasiparticles \cite{Shtengel,Feldman}.
In the manipulation of electron spins, a key element is the ability to induce transitions between the spin states and to prepare their arbitrary superpositions. This is commonly accomplished by magnetic resonance or via a local ac electric field \cite{Nowack,NPerge2,NPerge3,RLi,Kato,Rashba}. Here we will provide an alternative scheme to accomplish it by adjusting the orientation of the static magnetic field, which is in equivalent to simply exchanging two wavepacket-based non-Abelian quasiparticles.

The considered NQD system is a harmonically trapped electron with the Rashba-Dresselhaus coexisted SOC and subject to a static magnetic field. The governing two-component Schr\"odinger equation has the exact complete solution with different constants for the zero magnetic field case \cite{RLi} or under a modulation resonance condition with the corresponding strong magnetic field in the order of the harmonic trapping \cite{KHai,KHai1}. However, in the actual experiments \cite{NPerge3}, the magnetic field is much weaker than the harmonic trapping, which leads to that the exact solution is proven to be challenging. Treating the weak magnetic field as a perturbation, the perturbed solutions in series form have been constructed \cite{RLi} that leads to some interesting results, including a pathway for encoding the spin-orbit qubit and a method for determining
both the SOCs in the nanowire. It is worth noting that the normative quantum perturbation theory \cite{Landau,Zeng} supposed the energy eigenfunctions of the unperturbed equation as a complete orthogonal basic vector (in Hilbert space) which was applied to expand the corrected wave function as a series. Unfortunately, to avoid divergence of the series, the uncomplete orthogonal set lacking one basic vector was actually adopted. It has been proved that \cite{Zeng} if the missing component is added into the perturbed series, the corrected wave function may increase a phase factor which is unimportant for the uncoupled Schr\"odinger equation. But for the coupled two-component Schr\"odinger equation governing a two-spin system with SOC, such a phase factor may cause important coherence effect. Here we will apply a different quantum perturbation theory to avoid the loss of the coherence effect. In the previous works \cite{Hai1,Hai2}, we suggested an alternative theory for treating stationary-state quantum perturbation, which is based on the exact general solution of the single perturbed equation with arbitrary energy constants to be determined by the physical conditions called the boundedness conditions of the perturbed solutions. The general solution of a system can describe all properties of the system and more physics than any particular solution can. For a SO coupled system, the exact general solutions of two perturbed equations and the corresponding boundedness conditions warrant attention.
Given the general solution and the corresponding physical conditions, we are interested in how the wavepackets identical to the norms of the motional states replace the vortices in Refs. \cite{Stern,Kopnin} as the Majorana-like quasiparticles. Such quasiparticles behave as electroneutrality without Coulomb interaction between them, so that their interchange in one spatial dimension becomes possible with one wavepacket going through another without the classically impenetrable barrier \cite{KHai,KHai1,HChen}.

In this paper, we consider an electron confined in a gated NQD with Rashba-Dresselhaus coexisting SOC and weak static magnetic field, and treat the latter as a perturbation to seek the coherent superpositions of the two spin states, which just is the \emph{maximal spin-motion entangled states} \cite{Leibfried,Monroe,KHai2,HChen2}. For any unperturbed $n$ level with ground state $n = 0$ we obtain two coupled perturbed equations of the motinal states and their exact general solutions. From the \emph{boundedness and self-consistence conditions of the general solutions} we find two corrected energieshe with level splitting being much less than the unperturbed level difference, which correspond to a spin-orbit qubit. We demonstrate the metastability with high lifespan of the qubit by calculating the Einstein's spontaneous radiation coefficients \cite{Zeng}. This transparently reveals the qubit's insensitivity to the local perturbations and the weak noise from environment, and implies the \emph{decoherence-averse effect of SOC}. The \emph{quantization of phase-difference} between the two motinal states are found, which depends on the orientation of magnetic field for arbitrary fixed SOC. Thus the qubit can be coherently manipulated by adjusting the orientation of magnetic field. We show that such a quantized phase jump results in the spin flipping of the spin-motion entangled states and the position exchange of the probability-density wavepackets occupying the different spin states. The quantum quasiparticles described by the wavepackets can be proposed as the Majorana-like quasiparticles obeying the non-Abelian interchange after which the new state cannot be expressed as a product of the old state and a phase factor. Based on that the spin-orbit qubit can respond to both magnetic and electric fields \cite{RLi}, the results could be directly applied to an array of electrons separated from each other by different NQDs with weak neighboring coupling for coherently performing the robust quantum logic operations via the electric-magnetic combined modulations.

\section{Exact general solutions of the perturbed equations}

We consider a single electron confined in a gated NQD with Rashba-Dresselhaus coexisted SOC
and a 1D harmonic well controlled by the gate voltages on the static electric gates, and
subject to a weak static magnetic field \cite{Pershin,Nowak}. The Hamiltonian governing the system reads \cite{RLi}
\begin{eqnarray}
H&=&H_0+\alpha_D\sigma_x p+\alpha_R\sigma_y p+\frac 1 2 g(\sigma_x\cos \theta+\sigma_y\sin \theta), \nonumber\\
H_0&=&-\frac{1}{2}\frac{\partial^2}{\partial x^2}+\frac 1 2 x^2,
\end{eqnarray}
where we have adopted the natural unit system with $\hbar=m^*=\omega=1$ so that time, space and energy are in units of $\omega^{-1},\ L_h=\sqrt{\hbar/(m^*\omega)}$ and $\hbar\omega$. Here \cite{Isaacson} $m^*=0.014 m_e$ is the effective electron mass with $m_e$ the free-electron mass, $\omega$ denotes the harmonically trapped frequency, $\alpha_{R(D)}$ is the structure-related Rashba (bulk-originated Dresselhaus) SOC strength, $\sigma_{x(y)}$ is the $x (y)$ component of Pauli matrix, $g=g_e \mu_B B$ with $g_e$ denoting the gyromagnetic ratio \cite{Tsitsishvili}, $\mu_B$ being the Bohr magneton, $B$ and $\theta$ the strength and orientation of the static magnetic field. Applying the usual state vector $|\psi(t)\rangle=|\psi_{\uparrow} (t)\rangle|\uparrow \rangle+|\psi_{\downarrow} (t)\rangle|\downarrow \rangle$, the space-dependent state vector is defined as
\begin{eqnarray}
|\psi(x,t)\rangle=\langle x|\psi(t)\rangle=\psi_\uparrow (x,t)|\uparrow \rangle+\psi_\downarrow (x,t)|\downarrow \rangle
\end{eqnarray}
with $\psi_{\uparrow\downarrow}(x,t)=\langle x|\psi_{\uparrow\downarrow}(t)\rangle$ being the motional states entangling the corresponding spin states $|\uparrow \rangle=\left(\begin{array}{c}
  1 \\ 0
\end{array}\right)$ and $|\downarrow \rangle=\left(\begin{array}{cc}
  0 \\ 1
\end{array}\right)$, respectively. The normalization constants have been implied in the motional states. The spin-orbit entanglement of Eq. (2) requires the linear independencies \cite{KHai2,Kong} of the probability amplitudes $\psi_{\uparrow}(x,t)$ and $\psi_{\downarrow}(x,t)$. The probabilities of the particle occupying spin states $|\uparrow \rangle$ and $|\downarrow \rangle$ read $P_{\uparrow\downarrow}(t)= \int|\psi_{\uparrow\downarrow} (x,t)|^2dx$. The \emph{maximal spin-orbit entanglement} can be associated with \cite{KHai2} $P_{\uparrow}=P_{\downarrow}=\frac 1 2$. Applying Eqs. (1) and (2) to the Schr\"odinger equation $i\frac{\partial|\psi(x,t)\rangle}{\partial t}=H|\psi(x,t)\rangle$ and taking into account the stationary state solutions $\psi_{\uparrow\downarrow}(x,t)=e^{-iEt}\psi_{\uparrow\downarrow}(x)$ yields the coupled matrix equation \cite{KHai,KHai1}
\begin{eqnarray}
E\left(\begin{array}{c}  \psi_{\uparrow}(x) \\ \psi_{\downarrow}(x) \end{array}\right)&=&H_0\left(\begin{array}{c}
   \psi_{\uparrow} \\  \psi_{\downarrow} \end{array}\right)-i\alpha \frac{\partial}{\partial x}\left(\begin{array}{c}
    e^{-i\varphi}\psi_{\downarrow} \\  e^{i\varphi} \psi_{\uparrow} \end{array}\right)+\frac {g}{2}  \left(\begin{array}{c}  e^{-i\theta}\psi_{\downarrow} \\ e^{i\theta}\psi_{\uparrow} \end{array}\right), \nonumber\\
\alpha &=&\sqrt{\alpha_D^2+\alpha_R^2},\ \varphi=\arctan \frac{\alpha_R}{\alpha_D},
\end{eqnarray}
where we have used the definitions of the SOC strength and SOC-dependent phase as \cite{RLi} $\alpha\in [0,3]$ and $\varphi\in [0,\pi/2]$ for the Rashba-Dresselhaus SOC coexistence system.

Note that in the usual experiments \cite{Nowack,NPerge2,NPerge3}, the Zeeman splitting $(g\sim 1\mu$ eV) is much less than the orbit splitting $(\hbar\omega\sim 1$meV). Therefore, we can treat the term being proportional to $g$ of Eq. (3) as a perturbation. To decoupled the unperturbed terms, we make the function transformations
\begin{eqnarray}
\psi_{\uparrow\downarrow}(x)=\frac 1 2 e^{\mp i\varphi/2} [u(x)e^{-i\alpha x}\pm v(x)e^{i\alpha x}].
\end{eqnarray}
Here `$\uparrow$' corresponds to the above signs of `$\mp$' and `$\pm$', and `$\downarrow$' the below signs, respectively. Inserting Eq. (4) into Eq. (3), then multiplying the first line of the matrix equation by $e^{i(\varphi/2+\alpha x)}$ and multiplying the second line of the equation by $e^{-i(\varphi/2+\alpha x)}$, we obtain
\begin{widetext}
\begin{eqnarray}
E\left(\begin{array}{c} u \\ v \end{array}\right)&=&\Big(H_0-\frac{\alpha^2}{2}\Big)\left(\begin{array}{c}
   u \\  v \end{array}\right)+ \frac 1 2 g \left(\begin{array}{c}  e^{-i(\theta-\varphi)}(u-v e^{i 2\alpha x})+e^{i(\theta-\varphi)}(u+v e^{i 2\alpha x}) \\ e^{-i(\theta-\varphi)}(u e^{-i 2\alpha x}-v)-e^{i(\theta-\varphi)}(u e^{-i 2\alpha x}+v) \end{array}\right)\nonumber \\
&=&\Big(H_0-\frac{\alpha^2}{2}\Big)\left(\begin{array}{c}
   u \\  v \end{array}\right)+ \frac 1 2 g \left(\begin{array}{c} \cos(\theta-\varphi)u+i\sin(\theta-\varphi) e^{i 2\alpha x}v \\-i\sin(\theta-\varphi) e^{-i 2\alpha x}u -\cos(\theta-\varphi)v \end{array}\right).
\end{eqnarray}
\end{widetext}
In the SOC-magnetism phase-locked case \cite{KHai,KHai1} $\theta=\varphi$ and under a modulation resonance with the magnetic field strength being in order of the trap frequency, we have obtained the exact stationary-state solutions of Eq. (5). However, for arbitrary angle $\theta$ and strength $g$, the final term  of  Eq. (5) cannot be decoupled, so it is hard to construct an exact solution of the system. The corresponding perturbed solution of series form has been considered in Ref. \cite{RLi} by using the usual method \cite{Zeng}, which supplies a way to encode the spin-orbit qubits. Here we are interested in the exact general solutions of the first-order perturbed equations, by applying our different quantum perturbation theory \cite{Hai1,Hai2}. Thus we will find that there exist the boundedness and self-consistence conditions of the exact solutions which result in some new and interesting physics.

We adopt the Rayleigh-Schr\"odinger perturbation expansions up to any $j$th-order
\begin{eqnarray}
u(x)=\sum_{i=0}^j u^{(i)},\ \ v(x)=\sum_{i=0}^j v^{(i)},\ \ E=\sum_{i=0}^j E^{(i)}
\end{eqnarray}
for $j=1,2,\cdots$. Substituting Eq. (6) into Eq. (5) yields the decoupled zeroth-order equations and any $j$-th order perturbed equations
\begin{widetext}
\begin{eqnarray}
&&\Big(H_0-E^{(0)}-\frac{\alpha^2}{2}\Big)\left(\begin{array}{c}
   u^{(j)} \\  v^{(j)} \end{array}\right)=\left(\begin{array}{c}
   \epsilon^{(j)}_u \\ \epsilon^{(j)}_v \end{array}\right),\left(\begin{array}{c}
   \epsilon^{(0)}_u \\ \epsilon^{(0)}_v \end{array}\right)=\left(\begin{array}{c}
   0 \\0 \end{array}\right), \nonumber \\
&&\left(\begin{array}{c}
   \epsilon^{(j)}_u \\ \epsilon^{(j)}_v \end{array}\right)=\sum_{i=1}^jE^{(i)}\left(\begin{array}{c} u^{(j-i)} \\ v^{(j-i)} \end{array}\right)-\frac 1 2 g \left(\begin{array}{c} \cos(\theta-\varphi)u^{(j-1)}+i\sin(\theta-\varphi) e^{i 2\alpha x}v^{(j-1)} \\ -i\sin(\theta-\varphi) e^{-i 2\alpha x}u^{(j-1)} -\cos(\theta-\varphi)v^{(j-1)} \end{array}\right)\ \ \ \text{for} \ \ \ j=1,2,\cdots.
\end{eqnarray}
\end{widetext}
The zero-order equation contains two same harmonic oscillator equations with the well-known solutions \cite {RLi}
\begin{eqnarray}
u^{(0)}&=&u^{(0)}_n=C_n\psi_n(x),\ \ v^{(0)}=v^{(0)}_n=D_n\psi_n(x), \nonumber \\
E^{(0)}&=& E^{(0)}_n=\Big(\frac{1}{2}+n-\frac{\alpha^2}{2}\Big)
\end{eqnarray}
for $n=0,1,\cdots$. Here $\psi_n(x)$ denotes the well-known eigenstates of a harmonic oscillator, some real functions of $x$; $C_n$ and $D_n$ are the complex undetermined constants in the exponential forms with the phase difference $\phi$,
\begin{eqnarray}
C_n=|C_n|e^{i\phi_C},\ \ D_n=|D_n|e^{i\phi_D},\ \ \phi=\phi_C-\phi_D.
\end{eqnarray}
They will be determined by the normalization condition of the zero-order solution and the boundedness conditions of the first corrected solutions. The phase difference $\phi$ will bring important coherent effect. Given $E^{(j-1)}$, $u^{(j-1)}$ and $v^{(j-1)}$, any $j$th-order equation of Eq. (7) becomes an inhomogeneous linear ordinary differential equation. According to the constant variation method in ordinary differential equation theory, general solution of the inhomogeneous equation can be expressed as a sum of the general solution of the corresponding homogeneous equation and any particular solution of the inhomogeneous equation. Such a pair of exact general solutions can be expressed in teams of the integral forms \cite{Hai1,Hai2}
\begin{widetext}
\begin{eqnarray}
u^{(j)}_n&=& u^{(0)}_n\Big[A^{(j)}_{un}+\int_0^x \bar{u}^{(0)}_n\epsilon^{(j)}_{un} dx\Big]+\bar{u}^{(0)}_n\Big[B^{(j)}_{un}-\int_0^x u^{(0)}_n\epsilon^{(j)}_{un} dx\Big], \nonumber \\
v^{(j)}_n&=& v^{(0)}_n\Big[A^{(j)}_{vn}+\int_0^x \bar{v}^{(0)}_n\epsilon^{(j)}_{vn} dx\Big]+\bar{v}^{(0)}_n\Big[B^{(j)}_{vn}-\int_0^x v^{(0)}_n\epsilon^{(j)}_{vn} dx\Big], \ j=1,2,\cdots.
\end{eqnarray}
\end{widetext}
Here $(A^{(j)}_{un},B^{(j)}_{un})$ and $(A^{(j)}_{vn},B^{(j)}_{vn})$ are arbitrary constants determined by the physical conditions, namely the normalization and boundedness of the motional states, and the reality of the energy. The functions $\bar{u}^{(0)}_n= u^{(0)}_n\int [u^{(0)}_n]^{-2}dx$ and $\bar{v}^{(0)}_n= v^{(0)}_n\int [v^{(0)}_n]^{-2}dx$ are two unbounded solutions of the zeroth-order harmonic oscillator equations, respectively. The terms $A^{(j)}_{un}u^{(0)}_n+ B^{(j)}_{un}\bar{u}^{(0)}_n$ and $A^{(j)}_{vn}v^{(0)}_n+ B^{(j)}_{vn}\bar{v}^{(0)}_n$ are the general solutions of the two homogeneous equations in Eq. (7) with $j=0$. It has been proved that general solutions (10) are bounded if and only if the \emph{boundedness conditions} \cite{Hai1}
\begin{eqnarray}
I^{(j)}_{un\pm}&=&\lim_{x\to\pm \infty} \Big[B^{(j)}_{un}-\int_0^x u^{(0)}_n\epsilon^{(j)}_{un} dx\Big]=0, \nonumber \\
I^{(j)}_{vn\pm}&=&\lim_{x\to\pm \infty} \Big[B^{(j)}_{vn}-\int_0^x v^{(0)}_n\epsilon^{(j)}_{vn} dx \Big]=0
\end{eqnarray}
$(j=1,2,\cdots)$ are satisfied. Applying Eqs. (6) and (8) to Eq. (4) yields the probability amplitude
\begin{eqnarray}
\psi_{\uparrow\downarrow n}(x)&=&\sum_{j=0}^{\infty}\psi^{(j)}_{\uparrow\downarrow n}(x), \nonumber \\
\psi^{(j)}_{\uparrow\downarrow n}(x)&=& \frac 1 2 e^{\mp i \varphi/2} [u^{(j)}_ne^{-i\alpha x}\pm v^{(j)}_ne^{i\alpha x}].
\end{eqnarray}
The normalization conditions of the solution (2) are thereby \cite{Zeng}
\begin{eqnarray}
P_n&=& \int_{-\infty}^{\infty}\Big[\Big|\psi_{\uparrow n}(x)\Big|^2+\Big|\psi_{\downarrow n}(x)\Big|^2\Big]dx \nonumber \\
&=&\sum_{i=0}^{\infty}P^{(i)}_n=\sum_{i=0}^{\infty}[P^{(i)}_{\uparrow n}+P^{(i)}_{\downarrow n}]=1, \nonumber \\
P^{(0)}_n&=&1, \ P^{(i\ge 1)}_n=0,
\end{eqnarray}
where $P^{(i)}_{\uparrow\downarrow n}$ is a sum of the $i$th-order terms $O(g^i)$ in the integration $\int_{-\infty}^{\infty}\Big|\sum_{j=0}^{\infty}\psi^{(j)}_{\uparrow\downarrow n}(x)\Big|^2dx$.
The boundedness conditions (11) and the normalization conditions (13) determine the corrected energy $E^{(j)}_n$ and the normalization constants $C_n, D_n, A^{(j)}_{un},B^{(j)}_{un}$.
In the next section, we will focus on the first-order $(j=1)$ perturbed solutions to the ground state $(n=0)$ of the unperturbed zeroth-order equation. The similar treatment can be applied to any $j,n$ case.
\\

\section{Maximal spin-motion entangled states and energy corrections}

In the case $(j=0,n=0)$, combining Eqs. (8), (9) with Eq. (12), we get the zeroth-order and ground state vector
\begin{eqnarray}
|\psi^{(0)}_0(x,t)\rangle=e^{-iE^{(0)}_0 t}\Big[\psi^{(0)}_{\uparrow 0} (x)|\uparrow \rangle+\psi^{(0)}_{\downarrow 0}(x)|\downarrow \rangle\Big]
\end{eqnarray}
of Eq. (2) with the corresponding probability amplitudes and their norms being
\begin{eqnarray}
\psi^{(0)}_{\uparrow\downarrow0}(x)&=&\frac 1 2 e^{\mp i \varphi/2} [u^{(0)}_0e^{-i\alpha x}\pm v^{(0)}_0e^{i\alpha x}] \nonumber \\
&=&\frac 1 2 e^{i(\phi_C\mp\varphi/2)} [|C_0|e^{-i\alpha x}\pm |D_0|e^{i(\alpha x-\phi)}]\psi_0(x), \nonumber \\
|\psi^{(0)}_{\uparrow\downarrow0}(x)|^2&=&\frac 1 4 [|C_0|^2+|D_0|^2\pm 2 |C_0D_0|\cos(\phi-2\alpha x)] \nonumber \\
&&\times |\psi_0(x)|^2.
\end{eqnarray}
Clearly, the zeroth-order probability amplitudes $\psi^{(0)}_{\uparrow 0}(x)$ and $\psi^{(0)}_{\downarrow0}(x)$ are linearly independent for $\alpha\ne 0$. This means that the zeroth-order state (14) is a spin-motion entangled state \cite{Leibfried,Monroe,KHai2}, if and only if SOC exists. Such an entangled state just is the coherent superpositions of the two spin states. The final term of Eq. (15) describes the interference effect, which depends on the phase difference $\phi=\phi_C-\phi_D$ and the SOC strength $\alpha$. When the SOC vanishes, $\psi^{(0)}_{\uparrow 0}(x)$ and $\psi^{(0)}_{\downarrow0}(x)$ become linearly dependent, meaning existence of the \emph{decoherence-averse effect of SOC}. Thus we can suppress the decoherence \cite{Zou} by keeping the SOC. Applying the zeroth-order solutions (8) and the perturbation terms of Eq. (7), the first-order general solutions of Eq. (10) with $(j=1,n=0)$ become
\begin{widetext}
\begin{eqnarray}
u^{(1)}_0&=& u^{(0)}_0\Big[A^{(1)}_{u0}+\int_0^x \bar{u}^{(0)}_0\epsilon^{(1)}_{u0} dx\Big]+\bar{u}^{(0)}_0\Big[B^{(1)}_{u0}-\int_0^x u^{(0)}_0\epsilon^{(1)}_{u0} dx\Big], \nonumber \\
\epsilon^{(1)}_{u0}&=& E^{(1)}_0 u^{(0)}_0-\frac 1 2 g [\cos(\theta-\varphi)u^{(0)}_0+i\sin(\theta-\varphi) e^{i 2\alpha x}v^{(0)}_0], \nonumber \\
\bar{u}^{(0)}_0&=& u^{(0)}_0\int [u^{(0)}_0]^{-2}dx=C_0^{-1}\psi_0\int(\psi_0)^{-2}dx=\frac 1 2 C_0^{-1} \pi^{3/4}e^{-x^2/2} \text{Erfi}(x);  \nonumber \\
v^{(1)}_0&=& v^{(0)}_0\Big[A^{(1)}_{v0}+\int_0^x \bar{v}^{(0)}_0\epsilon^{(1)}_{v0} dx\Big]+\bar{v}^{(0)}_0\Big[B^{(1)}_{v0}-\int_0^x v^{(0)}_0\epsilon^{(1)}_{v0} dx\Big], \nonumber \\
\epsilon^{(1)}_{v0}&=& E^{(1)}_0 v^{(0)}_0+\frac 1 2 g [i\sin(\theta-\varphi) e^{-i 2\alpha x}u^{(0)}_0 +\cos(\theta-\varphi)v^{(0)}_0], \nonumber \\
\bar{v}^{(0)}_0&=& v^{(0)}_0\int [v^{(0)}_0]^{-2}dx=D_0^{-1}\psi_0\int(\psi_0)^{-2}dx=\frac 1 2 D_0^{-1} \pi^{3/4}e^{-x^2/2} \text{Erfi}(x).
\end{eqnarray}
\end{widetext}
Here we have used the ground state $\psi_0(x)=\pi^{-1/4}e^{-x^2/2}$ of a harmonic oscillator to derive the function $\text{Erfi}(x)=\text{Erf} (ix)/i$ with $\text{Erf}(ix)$ being the error function of $ix$. The solutions (8) and (16) should obey the normalization conditions $P^{(0)}_0=1, P^{(1)}_0=0$ of Eq. (13) and the boundedness conditions (11) for $j=1,n=0$. From them we can derive all the undetermined constants and the first-order corrected energy.

Knowing Eq. (15), the corresponding probabilities of the particle occupying spin states $|\uparrow \rangle$ and $|\downarrow \rangle$, and the normalization condition of the zero-order solution read
\begin{eqnarray}
P^{(0)}_{\uparrow\downarrow0}&=&\int_{-\infty}^{\infty}|\psi^{(0)}_{\uparrow\downarrow0}(x)|^2dx \nonumber \\
&=&\frac 1 4 [|C_0|^2+|D_0|^2\pm 2 |C_0D_0|\cos\phi \ e^{-\alpha^2}], \nonumber \\
P^{(0)}_0&=&P^{(0)}_{\uparrow0}+P^{(0)}_{\downarrow0}=\frac 1 2 [|C_0|^2+|D_0|^2]=1.
\end{eqnarray}
The latter gives a relation between $|C_0|$ and $|D_0|$. Inserting $\epsilon^{(1)}_{u0},\epsilon^{(1)}_{v0}$ of Eq. (16) and Eq. (8) into Eq. (11), from $I^{(1)}_{u0+}-I^{(1)}_{u0-}=0$ we get the first-order corrected energy
\begin{eqnarray}
E^{(1)}_0&=&\frac 1 2 g \int_{-\infty}^{\infty}u^{(0)}_0[\cos(\theta-\varphi)u^{(0)}_0 \nonumber \\
&&+i\sin(\theta-\varphi) e^{i 2\alpha x}v^{(0)}_0]dx \nonumber \\
&=&\frac 1 2 g \Big[\frac {iD_0}{C_0}\sin(\theta-\varphi) e^{-\alpha^2}+\cos(\theta-\varphi)\Big].
\end{eqnarray}
On the other hand, from $I^{(1)}_{v0+}-I^{(1)}_{v0-}=0$ we arrive at
\begin{eqnarray}
E^{(1)}_0&=&-\frac 1 2 g \int_{-\infty}^{\infty}v^{(0)}_0[i\sin(\theta-\varphi) e^{-i 2\alpha x}u^{(0)}_0 \nonumber \\
&&+\cos(\theta-\varphi)v^{(0)}_0]dx \nonumber \\
&=&\frac 1 2 g \Big[\frac {C_0}{iD_0}\sin(\theta-\varphi) e^{-\alpha^2}-\cos(\theta-\varphi)\Big].
\end{eqnarray}
Combining Eqs. (18) and (19) with Eq. (11), we obtain the two first order constants
\begin{eqnarray}
B^{(1)}_{u0}&=& \int_0^{\infty} u^{(0)}_0\epsilon^{(1)}_{u0} dx= \int_0^{-\infty} u^{(0)}_0\epsilon^{(1)}_{u0} dx  \nonumber \\
&=&-\frac {g}{\sqrt{\pi}} \sin(\theta-\varphi) \text{DawsonF}(\alpha),  \nonumber \\
B^{(1)}_{v0}&=&\int_0^{\infty} v^{(0)}_0\epsilon^{(1)}_{v0} dx=\int_0^{-\infty} v^{(0)}_0\epsilon^{(1)}_{v0} dx \nonumber \\
&=&\frac {g}{\sqrt{\pi}} \sin(\theta-\varphi) \text{DawsonF}(\alpha)
\end{eqnarray}
with $\text{DawsonF}(\alpha)=e^{-\alpha^2}\int_0^{\alpha}e^{y^2}dy$ being called the Dawson integration of $\alpha$. The other two first order constants $A^{(1)}_{u0}, A^{(1)}_{u0}$ can be determined by the first order normalization condition $P^{(1)}_0=0$ of Eq. (13), which are useful for computing the second-order corrected energy.

\begin{figure}[htp]
\includegraphics[height=1.53in,width=1.7in]{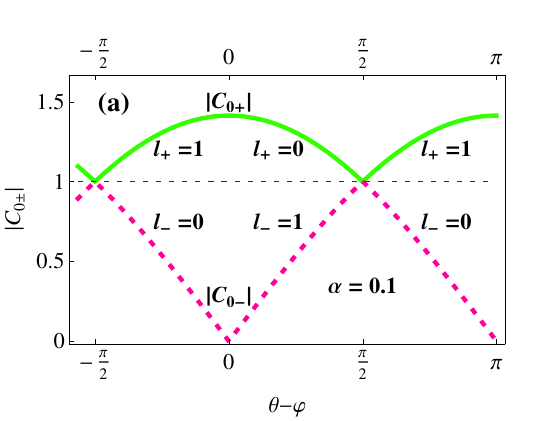}
\includegraphics[height=1.45in,width=1.65in]{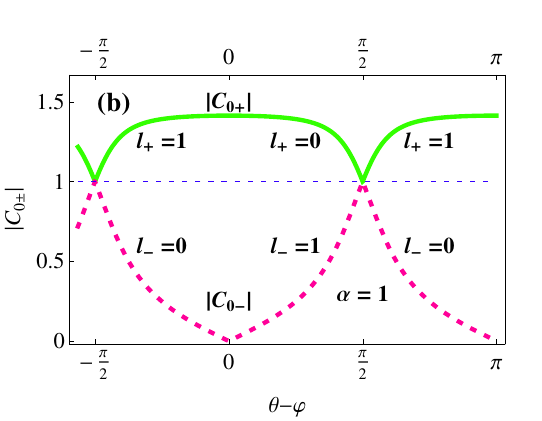}
\includegraphics[height=1.53in,width=1.7in]{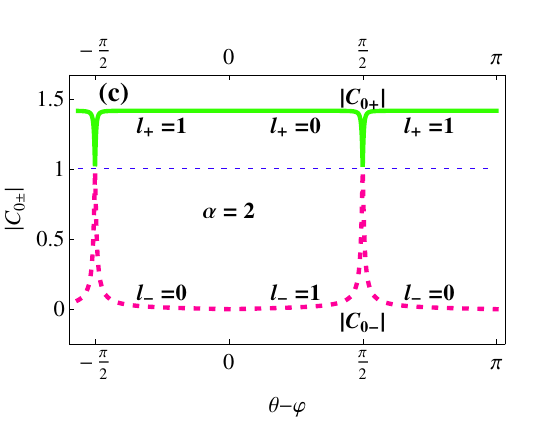}
\includegraphics[height=1.45in,width=1.65in]{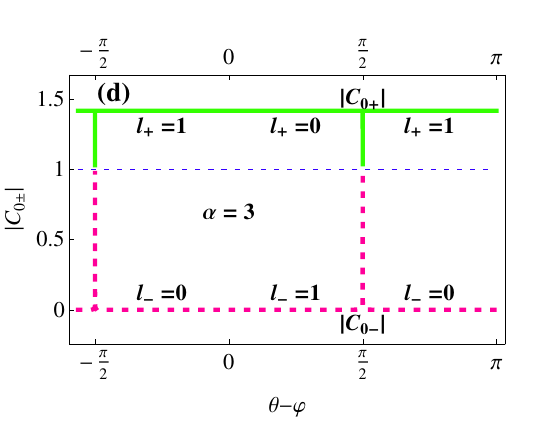}
\caption{(Color online)  Plots showing the relations (23) and (24) among the amplitudes $|C_{0+}|$ (solid curve), $|C_{0-}|$ (thick dashed curve),
phase differences $\phi=(l_{\pm}+1/2)\pi$ and system parameters $(\theta-\varphi)$ for $l_e=0, l_o=1, (\theta-\varphi)\in[-\pi/2,\pi]$ and (a) $\alpha= 0.1$, (b) $\alpha= 1$, (c) $\alpha= 2$, (d) $\alpha= 3$. The thin dashed line indicates the touch value $|C_{0+}|=|C_{0-}|=1$. All the quantities plotted in the figures of this paper are dimensionless.}
\end{figure}

It is important for us to derive the energy correction $E^{(1)}_0$ from Eqs. (18) and (19). At first we notice that the energy must be a real constant that requires the undetermined constant $\frac {C_0}{iD_0}=\frac {|C_0|}{|D_0|}e^{i(\phi-\pi/2)}$ is a real constant. This gives the \emph{quantized phase difference}
\begin{eqnarray}
\phi=\Big(l+\frac 1 2\Big)\pi \ \ \ \text{for}\ \ \ \emph{l}=0,1,2\cdots.
\end{eqnarray}
We will call $l$  the phase quantum number. By such a phase difference we mean that Eq. (17) gives $P^{(0)}_{\uparrow0}=P^{(0)}_{\downarrow0}=\frac 1 2$ and Eq. (14) is \emph{the maximal spin-orbit entangled state}. Then let $E^{(1)}_0$ of Eq. (18) be the same with that of  Eq. (19) we derive the \emph{self-consistence conditions} of the general solutions (16) as
\begin{eqnarray}
2\cos(\theta-\varphi)&=&e^{il\pi}\sin(\theta-\varphi) e^{-\alpha^2}\Big[\frac {|C_0|}{|D_0|}-\frac {|D_0|}{|C_0|}\Big], \nonumber \\
\tan(\theta-\varphi)&=&e^{il\pi+\alpha^2}\frac {2|C_0D_0|}{|C_0|^2-|D_0|^2} \nonumber \\
&=&e^{il\pi+\alpha^2}\frac {|C_0|\sqrt{2-|C_0|^2}}{|C_0|^2-1}.
\end{eqnarray}
In the calculation, $e^{-il\pi}=e^{il\pi}$ and Eq. (17) have been adopted. Given Eqs. (22) and (17), the simple calculation yields
\begin{eqnarray}
|C_0|&=&|C_{0\pm}|=\sqrt {1\pm [1+\tan^2(\theta-\varphi)e^{-2\alpha^2}]^{-1/2}}, \nonumber \\
|D_0|&=&|D_{0\pm}|=\sqrt {1\mp [1+\tan^2(\theta-\varphi)e^{-2\alpha^2}]^{-1/2}}.
\end{eqnarray}
They confine the regions of $|C_{0\pm}|$ and $|D_{0\pm}|$ as $|C_{0-}|=|D_{0+}| \in[0,1],\ |C_{0+}|=|D_{0-}|\in[1, \sqrt 2]$ and directly affect the state (14) and energy correction (18) and (19). At $\tan(\theta-\varphi)=0$, we have $|C_{0-}|=0,|C_{0+}|=\sqrt 2$, and $\tan(\theta-\varphi)=\pm\infty$ means $|C_{0-}|=|C_{0+}|=1$ with signs $\pm$ depending on different $l=l_{\pm}$. The self-consistence conditions (22) reveal several important relations between the undetermined constants $(|C_{0\pm}|,l_{\pm})$ and the system parameters $(\theta,\varphi,)$:
\begin{eqnarray}
&&\infty > \tan(\theta-\varphi)\ge 0, \ \Big \{ \begin{array}{c}
  |C_{0}|= |C_{0-}|, \ l=l_-=l_o, \\  |C_{0}|= |C_{0+}|,\ l=l_+=l_e;
\end{array}  \nonumber \\
&&-\infty < \tan(\theta-\varphi)\le 0, \ \Big \{ \begin{array}{c}
  |C_{0}|= |C_{0-}|, \ l=l_-=l_e, \\  |C_{0}|= |C_{0+}|,\ l=l_+=l_o.
\end{array} \ \ \
\end{eqnarray}
The signs $l_e$ and $l_o$ denote the even and odd numbers respectively, so that we have $e^{il_e\pi}=-e^{il_o\pi}=1$. Equations (24) and (21) reveal that \textbf{\emph{the sign-changing points of $\tan(\theta-\varphi)$, $(\theta-\varphi)=\pm k \pi/2$ with $k =0, 1, \cdots$, just are the quantized phase jump points at which phase difference $\phi$ jumps between $l_\pm=l_e$ and $l_\pm=l_o$.
\begin{figure}[htp]
\includegraphics[height=1.53in,width=1.7in]{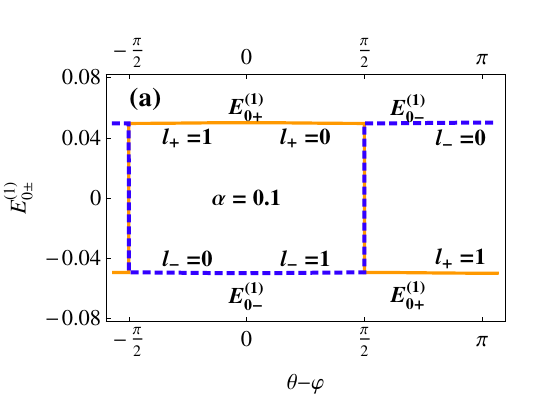}
\includegraphics[height=1.45in,width=1.65in]{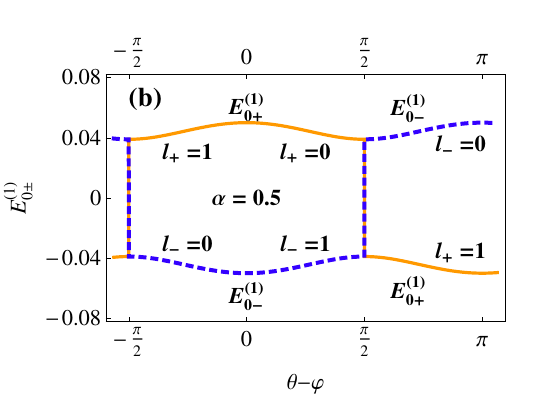}
\includegraphics[height=1.53in,width=1.7in]{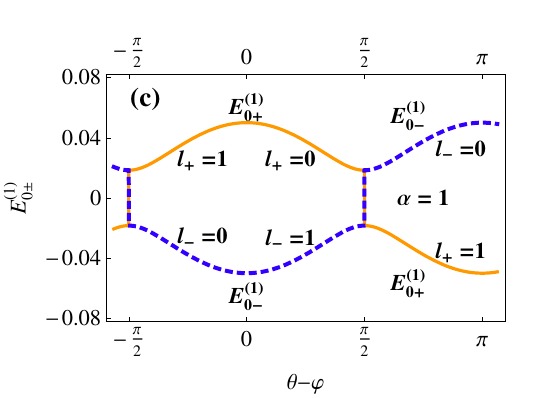}
\includegraphics[height=1.45in,width=1.65in]{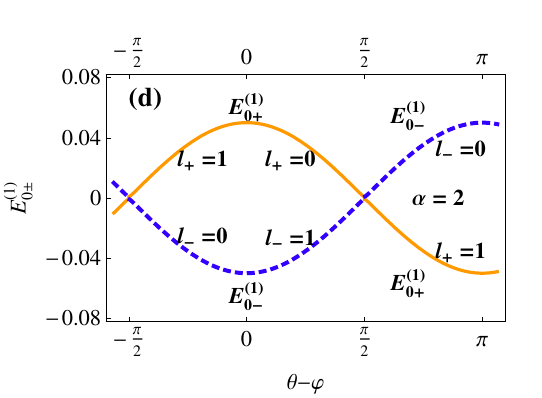}
\caption{(Color online)  Plots of the first-order corrected energies $E^{(1)}_{0+}$ (solid curve) and $E^{(1)}_{0-}$ (dashed curve) of Eq. (25) as functions of
$(\theta-\varphi)$ for $g=0.1$ and (a) $\alpha=0.1$, (b) $\alpha=0.5$, (c) $\alpha=1$, (d) $\alpha=2$. The phase quantum numbers $l_{\pm}$ are labeled according to Eq. (24) and the correspondence between $E^{(1)}_{0\pm}$ and $|C_{0\pm}|$. }
\end{figure}
Such phase jumps can be manipulated by adjusting the magnetic field angle  $\theta$ or SOC phase $\varphi$, and consequently cause the transitions of states (15) and (14).}} The relations (23) and (24) are clearly shown by the figure $|C_{0\pm}|$ vs $(\theta-\varphi)$ as in Fig. 1. In this figure we also show that the maximal value $|C_{0+}|=\sqrt2$ and minimal value $|C_{0-}|=0$ appear at $(\theta-\varphi)=k\pi$ for $k = 0,\pm1$ and any $\alpha$ value. With the increase of  $\alpha$ value, amplitudes $|C_{0\pm}|$ tend to two straight lines, approximately, $|C_{0+}|=\sqrt2$ and $|C_{0-}|=0$,
except for the special value $|C_{0\pm}|=1$ at the point $(\theta-\varphi)=\pm \pi/2$.

Applying  Eqs. (21-24)  to  Eq. (18)  produces  the  corrected  energy
\begin{eqnarray}
E^{(1)}_{0}&=&E^{(1)}_{0\pm}=E^{(1)}_{0}(|C_{0\pm}|,l_\pm) \nonumber \\
&=&\frac g 2 \cos(\theta-\varphi)\Big[1-e^{il_\pm\pi-\alpha^2}\frac {|D_{0\pm}|}{|C_{0\pm}|}\tan(\theta-\varphi)\Big] \nonumber \\
&=&\pm\frac g 2 \cos(\theta-\varphi)\sqrt{1+\tan^2(\theta-\varphi)e^{-2\alpha^2}}.
\end{eqnarray}
The level splitting $|E^{(1)}_{0+}-E^{(1)}_{0-}|=2|E^{(1)}_{0+}|=2|E^{(1)}_{0-}|=g\sqrt{\cos^2(\theta-\varphi)+\sin^2(\theta-\varphi)e^{-2\alpha^2}}$ is in agreement with that of Ref. \cite{RLi}. We plot the first-order corrected energies as functions of $(\theta-\varphi)$ for several different $\alpha$ values in Fig. 2. From this figure we can see that in the interval $(\theta-\varphi)\in(-\pi/2,\pi/2)$ corresponding to $\cos(\theta-\varphi)>0$, it is shown that $E^{(1)}_{0+}>0$ and $E^{(1)}_{0-}<0$. While in the interval $(\theta-\varphi)\in(\pi/2,\pi)$ associated with $\cos(\theta-\varphi)<0$, we observe that $E^{(1)}_{0+}<0$ and $E^{(1)}_{0-}>0$. At the phase jump points $(\theta-\varphi)=\pm k\pi/2$, the corrected energy may be continues for an even number $k$. While for an odd $k$ the energy inversions may be found, where $E^{(1)}_{0\pm}$ inverts and hops a height $ge^{-\alpha^2}$, and tends to $0$ with the increase of $\alpha$ value that leads to the approximate \emph{level crossing}. For any fixed $\alpha$ value, Fig. 2 shows
the maximal absolute value $|E^{(1)}_{0\pm}|$ at $(\theta-\varphi) =0, \pi$, and the minimal $|E^{(1)}_{0\pm}|$ at $(\theta-\varphi) =\pm \pi/2$.

In Figs. 1 and 2 we have observed that for a set of experimentally tunable system parameters $(g,\theta,\varphi,\alpha)$, there exist two sets $S_+$ and $S_-$ of constants, $S_{\pm}=\{|C_{0\pm}|,|D_{0\pm}|,l_{\pm},E^{(1)}_{0\pm}\}$ associated with the signs $``+"$ and $``-"$ respectively. While any constant set corresponds to a state vector of Eq. (14) and energy correction of Eq. (25). Hereafter, we will consider the states to leading order and the energies to first order. Adopting the phase quantum number $l_{\pm}$ to label the states and using Eqs. (21) and (23), we rewrite the maximal spin-motion entangled states (14) as the space-dependent state vectors
\begin{eqnarray}
&&|\psi^{(0)}_{0l_{\pm}}(x,t)\rangle=e^{-iE_{0\pm} t}\Big[\psi^{(0)}_{\uparrow 0l_{\pm}} (x)|\uparrow \rangle+\psi^{(0)}_{\downarrow 0l_{\pm}}(x)|\downarrow \rangle\Big],\\
&&\psi^{(0)}_{\uparrow0l_{\pm}}=\frac 1 2e^{i(\phi_C\mp\varphi/2)} \Big[|C_{0\pm}|e^{-i\alpha x}+ |D_{0\pm}|e^{i[\alpha x-(l_{\pm}+\frac 1 2)\pi]}\Big]\psi_0, \nonumber \\
&&\psi^{(0)}_{\downarrow0l_{\pm}}=\frac 1 2e^{i(\phi_C\mp\varphi/2)} \Big[|C_{0\pm}|e^{-i\alpha x}- |D_{0\pm}|e^{i[\alpha x-(l_{\pm}+\frac 1 2)\pi]}\Big]\psi_0. \nonumber
\end{eqnarray}
Note that Eq. (26) contains the four state vectors: $|\psi^{(0)}_{0,l_+=l_e}\rangle,\ |\psi^{(0)}_{0,l_+=l_o}\rangle,\ |\psi^{(0)}_{0,l_-=l_e}\rangle$ and $|\psi^{(0)}_{0,l_-=l_o}\rangle$ associated with two different $(\theta-\varphi)$ values. For a single fixed $(\theta-\varphi)$, $l_+$ and $l_-$ are fixed that means Eq. (26) including only two states with $l_+\ne l_-$ and $E_{0+}\ne E_{0-}$. Obviously, at the phase jump points, the changes of $l_\pm$ between $l_e$ and $l_o$ result in the motional-state exchanges between $\psi^{(0)}_{\uparrow0l_{\pm}}(x)$ and $\psi^{(0)}_{\downarrow0l_{\pm}}(x)$. This is equivalent to the \emph{spin flipping} in the spin-motion entangled states  $|\psi^{(0)}_{0l_{\pm}}(x,t)\rangle$.

Now we simply prove the metastability of the stationary states by calculating the Einstein's spontaneous radiation coefficients \cite{Zeng}. Taking into account the periodic weak noise from environment which includes the ac electric potential $\chi x\cos(\omega_{\pm} t)$ with small strength $\chi$ and frequency $\omega_{\pm}=|E^{(1)}_{0+}-E^{(1)}_{0-}|/\hbar=2|E^{(1)}_{0\pm}|/\hbar$ given by Eq. (25). After a long and easy calculation, the spontaneous radiation coefficient of the electron from the positive-energy $|E^{(1)}_{0\pm}|$ state to the negative-energy $-|E^{(1)}_{0\pm}|$ state is obtained as \cite{Zeng}
\begin{eqnarray}
A_{sr}&=&\frac {4e^2\omega^3_{\pm}}{3\hbar c^3}|x_{l_+,l_-}|^2 \nonumber \\
&=&\frac {4e^2\omega_{\pm}^3}{3\hbar c^3}|\langle\psi^{(0)}_{0l_+}(x,t)|x|\psi^{(0)}_{0l_-}(x,t)\rangle|^2 \nonumber \\
&=&\frac {4e^2\omega_{\pm}^3\alpha^2\sin^2\varphi}{3\hbar c^3[e^{2\alpha^2}+\tan^2(\theta-\varphi)]}.
\end{eqnarray}
In the calculation, Eqs. (26), (23) and $l_+=0, l_-=1$ have been used. At the phase jump points $(\theta-\varphi)=\pm k \pi/2$, we have the minimal spontaneous radiation coefficient $A_{srmin}=0$ for an odd $k$, which corresponds to infinite lifespan of a stable qubit. We also obtain the maximal spontaneous radiation coefficient
\begin{eqnarray}
A_{srmax}=\frac {4e^2\omega_{\pm}^3\alpha^2\sin^2\varphi}{3\hbar c^3e^{2\alpha^2}}\sim 10^{-6}/s
\end{eqnarray}
for an even $k$, where we have taken \cite{RLi} $g/\hbar= 0.03 \omega$ in order of $10^{10}$Hz and $\alpha^2e^{-2\alpha^2}\sin^2\varphi\sim 0.1$.
The corresponding minimal lifetime $1/A_{srmax}$ of the positive-energy state is in order of Ms, a longer time for the microcosmic system. For any other $(\theta-\varphi)$ value, the corresponding lifespan is greater than the minimal lifetime. While the negative-energy state is a ground state with lifespan being much greater than $1/A_{srmax}$. So \emph{we can regard these states as metastable ones and the related qutbit is robust}. This transparently reveals the qubit's insensitivity to the local perturbations and the weak noise from environment. In addition, both the level spacing and the spontaneous radiation lifespan have periodic responses to the direction of the static magnetic field. These responses can be used to determine the magnitude $\alpha$ and phase $\varphi$ of SOC in the nanowire \cite{RLi}.
The two metastable states $|\psi^{(0)}_{0l_{+}}(x,t)\rangle$ and $|\psi^{(0)}_{0l_{-}}(x,t)\rangle$ are associated with two levels $E_{0\pm}$ and one spin-orbit qubit. Because at any moment, the electron can be in only one state of them. Consequently, we can apply an external ac electric field to the system to make the resonance of the driving frequency and the level splitting of the spin-orbit qubit such that the transitions happen between the two states \cite{RLi,KHai,KHai1}. The level splitting is much less than the zero-order quantum gap $\hbar \omega$, so the resonance transition is insensitive to the environment, and the coherence of the two-level system can be kept well.

It is worth noting that the key element of spin manipulation is the ability to induce transitions between the spin states and to prepare their arbitrary superpositions \cite{Nowack,NPerge2,NPerge3,RLi,Kato,Rashba}. In next section, we will demonstrate that this can be accomplished in a transparent manner, namely by using the orbital part of the spin-motion entangled states (26) to adjust the orientation angle of the static magnetic field for changing the $(\theta-\varphi)$ value, which is in equivalent to the spin flipping and the wavepacket-based non-Abelian quasiparticles exchanging.

\begin{figure*}[htp]
\includegraphics[height=2.0in,width=2.55in]{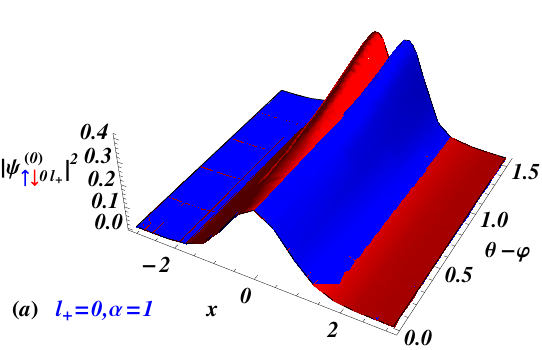}\ \ \
\includegraphics[height=2.0in,width=2.55in]{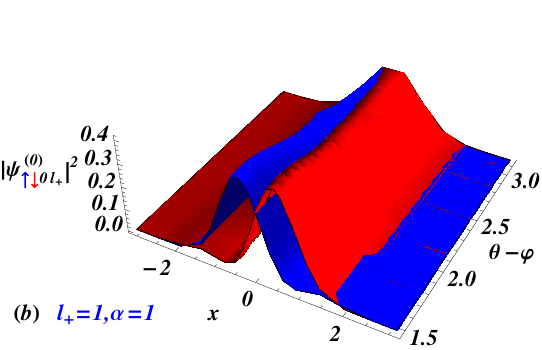}
\includegraphics[height=2.0in,width=2.55in]{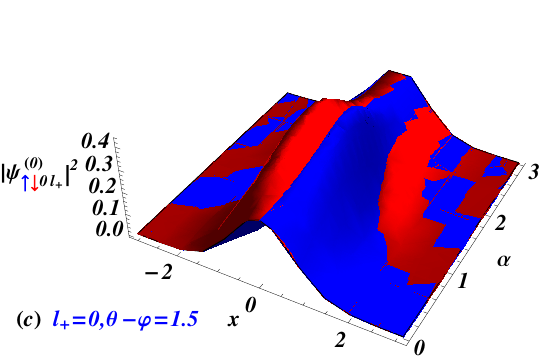}\ \ \
\includegraphics[height=2.0in,width=2.55in]{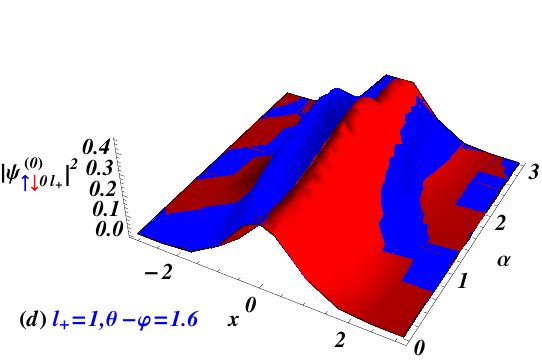}
\caption{(Color online) The 3D wavepackets of the probability densities (29) for (a) $l_+ = 0, \alpha =1$ and varying $(\theta-\varphi)$ from $0$ to $\pi/2$; (b) $l_+ = 1, \alpha =1$ and varying $(\theta-\varphi)$ from $\pi/2$ to $\pi$, (c) $l_+ = 0, (\theta-\varphi)=1.5<\pi/2$, and varying $\alpha$ from $0$ to $3$; (d) $l_+ = 1, (\theta-\varphi)=1.6>\pi/2$, and varying $\alpha$ from $0$ to $3$. The blue wavepackets describe $|\psi^{(0)}_{\uparrow0l_{+}}|^2$ and the red means $|\psi^{(0)}_{\downarrow0l_{+}}|^2$.}
\end{figure*}

\section{Coherently controlling qubits by exchanging the non-Abelian quasiparticles}

Given the maximal spin-motion entangled states (26), we are interested in how the density wavepackets identical to the norms of the motional states replace the vortices in Refs. \cite{Stern,Kopnin} as the Majorana-like quasiparticles obeying non-Abelian exchang. Such quasiparticles behave as electroneutrality without Coulomb interaction between them, so that their interchange in one spatial dimension becomes possible with one wavepacket going through another without the classically impenetrable barrier \cite{KHai,KHai1,HChen}.
Noticing $\cos \phi=\cos(l_{\pm}+\frac 1 2)\pi=0,\sin \phi=\cos(l_{\pm}\pi)$ and Eq. (17), the probability densities of Eq. (15) occupying different spin states (26) become
\begin{eqnarray}
|\psi^{(0)}_{\uparrow0l_{\pm}}|^2&=&\frac 1 2 [1+ |C_{0\pm}D_{0\pm}|\cos(l_{\pm}\pi) \sin(2\alpha x)]|\psi_0|^2,\nonumber \\
|\psi^{(0)}_{\downarrow0l_{\pm}}|^2&=&\frac 1 2 [1- |C_{0\pm}D_{0\pm}|\cos(l_{\pm}\pi) \sin(2\alpha x)]|\psi_0|^2.\ \ \ \
\end{eqnarray}
By $l_{\pm}$ we mean the corresponding constant set $S_{\pm}=\{|C_{0\pm}|,|D_{0\pm}|,l_{\pm},E^{(1)}_{0\pm}\}$ is used, respectively for the signs $``+"$ and $``-"$.
Equation (29) and Fig. 2 tell us that for a small variation from the phase jump points $(\theta-\varphi)=\pm k \pi/2$ the densities exchange between $|\psi^{(0)}_{\uparrow0l_{\pm}}|^2$ and $|\psi^{(0)}_{\downarrow0l_{\pm}}|^2$.
In order to see the exchanges of quasiparticles, from Eqs. (29), (23) and (24) we plot four 3D figures for displaying evolution of the density wavepackets with the coordinate $x$ and the parameter $(\theta-\varphi)$ and $\alpha$, as shown in Fig. 3. In these figures, we concentrate on the zeroth order probability densities, because the perturbation corrections to them are ignorable \cite{Hai2}. In Figs. 3(a) and 3(b), we show that the wavepackets exchange positions and arrive at the largest distance between them at $(\theta-\varphi)=\pi/2$ and for $\alpha\sim 1$. They close to each other with the increase or decrease of $(\theta-\varphi)$ from $\pi/2$ to $\pi$ or to $0$, where the minimal distance zero reaches. By Figs. 3(c) and 3(d), we display evolution of the wavepackets with SOC strength $\alpha$ before and after the position exchange, respectively. The distance between wavepackets change with $\alpha$, and their largest distance appears at $\alpha\approx 1$ for $(\theta-\varphi)=1.5<\pi/2$ and for $(\theta-\varphi)=1.6>\pi/2$. From Eqs. (23) and (24) we observe $|C_{0+}D_{0+}|=|C_{0-}D_{0-}|$ and $l_+\ne l_-$ implying $\cos(l_{+}\pi)=-\cos(l_{-}\pi)$ such that Eq. (29) means $|\psi^{(0)}_{\uparrow0l_{-}}|^2=|\psi^{(0)}_{\downarrow 0l_{+}}|^2$ and $|\psi^{(0)}_{\downarrow 0l_{-}}|^2=|\psi^{(0)}_{\uparrow0l_{+}}|^2$. Therefore, we take only $|\psi^{(0)}_{\uparrow0l_{+}}|^2$ and $|\psi^{(0)}_{\downarrow0l_{+}}|^2$ as examples in Figs. 3 and 4.

We now illustrate that such exchanges of quasiparticles based on probability densities correspond to the state transitions between $|\psi^{(0)}_{0l_{\pm}=0}(x,t)\rangle$ and $|\psi^{(0)}_{0l_{\pm}=1}(x,t)\rangle$ with the spin flipping. They can be divided into two cases according Fig. 2, Fig. 3 and Eq. (26) as the following.

Case 1. Exchanging positions of two small-distance wavepackets with varying $(\theta-\varphi)$ from $(k -\lambda) \pi$ to $(k +\lambda) \pi$ for $0<\lambda \ll 1$, as shown in Figs. 4(a) and 4(b) with  $k=0, \lambda=0.1$. Such an exchange changes the states and shifts the energy of Fig. 2 only an ignorable value. Such operations of quantum states are insensitive to the environment, and similar to the topological quantum operations of the degenerate ground states without level difference \cite{KHai}.

\begin{figure}[htp]
\includegraphics[height=1.53in,width=1.7in]{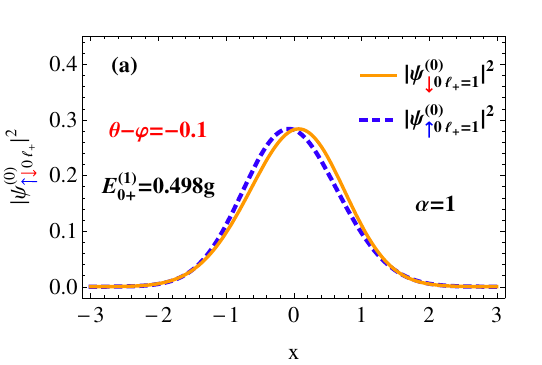}
\includegraphics[height=1.45in,width=1.65in]{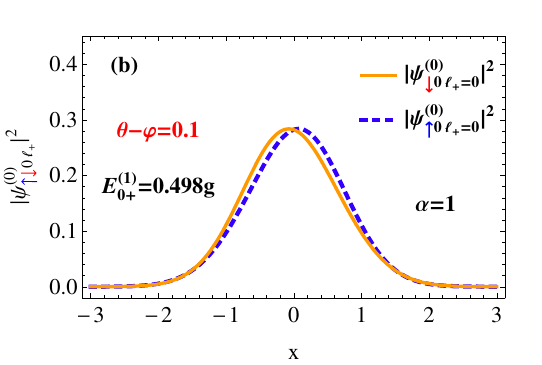}
\includegraphics[height=1.53in,width=1.7in]{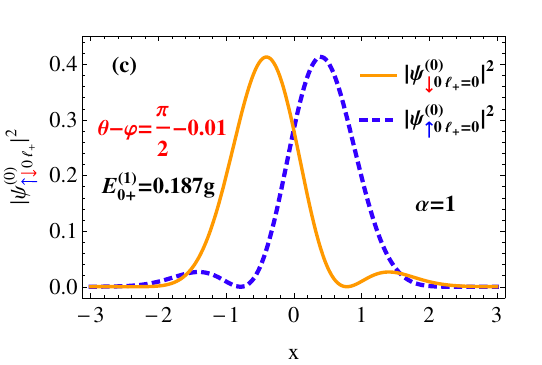}
\includegraphics[height=1.45in,width=1.65in]{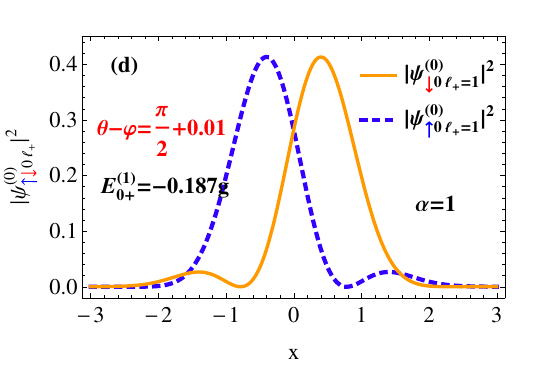}
\caption{(Color online)  Plots of the spatial evolutions of the quasiparticle wavepackets $|\psi^{(0)}_{\uparrow0l_{+}}|^2$ (dashed curve) and $|\psi^{(0)}_{\downarrow0l_{+}}|^2$ (solid curve) for $\alpha=1$ and (a) $(\theta-\varphi)=-0.1, l_+ =1, E^{(1)}_{0+}=0.498g$; (b) $(\theta-\varphi)=0.1, l_+ = 0, E^{(1)}_{0+}=0.498g$; (c) $(\theta-\varphi)=\frac {\pi}{2}-0.01, l_+ =0, E^{(1)}_{0+}=0.187g$; (d) $(\theta-\varphi)=\frac {\pi}{2}+0.01, l_+ =1, E^{(1)}_{0+}=-0.187g$. The position exchange and energy keep of the quasiparticles at $(\theta-\varphi)=0$ are shown by 4(a) and 4(b), and the position exchange and energy inversion at $(\theta-\varphi)=\frac {\pi}{2}$ are exhibited by 4(c) and 4(d).}
\end{figure}

Case 2. Exchanging positions of two large-distance wavepackets with adjusting $(\theta-\varphi)$ from $(k+\frac 1 2 -\lambda) \pi$ to $(k+\frac 1 2 +\lambda) \pi$ for $0<\lambda \ll 1$, as shown in Figs. 4(c) and 4(d) with  $k=0, \lambda=0.01$. Such an exchange transfers the states and inverts the energy between $E^{(1)}_{0+}$ and $-E^{(1)}_{0+}$. In the two figures, we observe the distance between peaks of wavepackets being about $\langle\psi^{(0)}_{0l_+=0}(x,t)|x|\psi^{(0)}_{0l_+=1}(x,t)\rangle\approx 2(L_h)=2[\sqrt{\hbar/(m^*\omega)}]\sim 10$nm.

The exchanges of wavepacket positions are performed through the changes of the coherent terms of Eq. (29). So we can identify quantum interferences as the heart of the above quantum control, and call the operation of the density exchanges the \emph{coherent manipulation} \cite{Fielding,KXiao}. Particularly, the quantum quasiparticles described by such wavepackets obey the non-Abelian interchange after which the new state of Eq. (26) cannot be expressed as a product of the old state and a phase factor, and can be proposed as the Majorana-like quasiparticles thereby. Under some given initial conditions, motions of the quasiparticles possess the 1D helicity \cite{NPerge3}, namely the moved direction of the spin-up quasiparticle differs from that of the spin-down quasiparticle. The exchanges of quasiparticles imply some unclear spatiotemporal evolutions similar to the resonance transitions, which may also be quantum-mechanically allowable \cite{Zeng}.

\section{Conclusion and discussion}

We have investigated a single spin-orbit coupled electron confined in a gated NQD with Rashba-Dresselhaus coexisted SOC and weak static magnetic field. Treating the weak field as a perturbation, we obtain the maximal spin-motion entangled states with the exact general solutions of the perturbed equations. We find that there exist two corrected energies to any level of the unperturbed system for fitting the boundedness and self-consistence conditions of the general solutions. The level splitting are much less than the unperturbed level difference and the perturbed state corresponds to a spin-orbit qubit. We calculate the Einstein's spontaneous radiation coefficients \cite{Zeng} by which we reveal the qubit's insensitivity to the local perturbations and the weak noise from environment. The quantized phase-difference between the two motinal states are found, which depend on the orientation angle of magnetic field for arbitrary fixed SOC. Thus the qubit can be coherently manipulated by adjusting the orientation of magnetic field. We show that such a quantized phase jump results in the spin flipping of the spin-motion entangled states and the position exchange of the probability-density wavepackets occupying the different spin states. The quantum quasiparticles described by the wavepackets can be proposed as the Majorana-like quasiparticles obeying the non-Abelian interchange. The operations based on the interchanges of the non-Abelian quasiparticles may be robust . The spin-motion entanglement depend on the existence of SOC that implies the decoherence-averse effect of SOC. While the operations of quasiparticle exchanges depend on the interference terms of the probability-density wavepacket, meaning the coherent controls.

For a fixing $(\theta-\varphi)$ value, Eq. (26) includes two states with $l_+\ne l_-$ and $E_{0+}\ne E_{0-}$. From Fig. 2 we have seen that $E_{0+}=- E_{0-}>0$ for $|\theta-\varphi|<\frac {\pi}{2}$, and $E_{0+}=- E_{0-}<0$ for $|\theta-\varphi|>\frac {\pi}{2}$. We can initially prepare a ground state with the lower energy and create a quantum transition from the ground state to the excitation state with higher energy, by using a laser with resonance frequency to match the level difference \cite{RLi} $|E_{0+}- E_{0-}|$.  Generally, the usual quantum transition with energy exchange is equivalent to the state transfer. Differing from the resonance transition, we create the quantum transition in this work by fine tuning the orientation angle $\theta$ of magnetic field to vary $(\theta-\varphi)$ value for fixed other experimental parameters. Treating the maximal spin-motion entangled states (26) as leading-order solutions, the obtained results could be transparently applied to an array of electrons separated from each other by different quantum dots with weak neighboring coupling as another perturbation. Although the operation to angle $\theta$ of the static magnetic field cannot be performed individually for the electrons in quantum-dot-array, it is useful for combining with a local ac electric field to create the required initial state, then to perform the individual operation to any one of qubits by using the electric-magnetic combined modulations \cite{Nowack,RLi,Loss,KHai}, which could be fundamental important for encoding the robust qubits and accomplishing the spin-based quantum information processing.

In addition, because the two sets of time-independent motional states are associated with different energies $E^{(1)}_{0\pm}$, they obey two different stationary-state equations of Eq. (3) for $E^{(1)}_{0+}$ and $E^{(1)}_{0-}$, respectively. Therefore, their linear superposition is not a solution of Eq. (3). However, the linear superposition of the two time-dependent state vectors (26) still is a solution of the time-dependent Schr\"odinger equation. Such a superposition state denotes a nonstationary coherent state with periodically variable probability amplitudes. Their physical properties and possible applications in coherent manipulation of electron spins should be further investigated.
It is also quite interested to apply our method for seeking quantum chaos not only for the spatiotemporal coordinate but also for spin in a NQD electron system with SOC and driven multiminima potential \cite{ESherman}. The results will lead to a new method of the chaotic spin-manipulation in nanostructures.

\bf Acknowledgments \rm
This work was supported by the NNSF of China under Grant Nos. 12375012 and 12247105, the Hunan Provincial Major Sci-Tech Program under Grant No. 2023zk1010, XJ2302001, and the Scientific Research Fund of Hunan Provincial Education Department of China under Grant No. 22A0032.

\end{document}